\documentstyle[12pt,epsf]{article} 
\newcommand{\GeV}{\mbox{ GeV}}
\newcommand{\TeV}{\mbox{ TeV}}
\newcommand{\pb}{\mbox{ pb}}
\newcommand{\ie}{{i.e. }}
\newcommand{\slE}{\mbox{/}\!\!\!\!E}
\catcode`@=11 
\def\lsim{\mathrel{\mathpalette\@versim<}}
\def\gsim{\mathrel{\mathpalette\@versim>}}
\def\gtlt{{\smash{\lower0.4ex\hbox{$<$}} \atop \smash{\raise0.4ex\hbox{$>$}}}}
\def\@versim#1#2{\lower0.2ex\vbox{\baselineskip\z@skip\lineskip\z@skip
  \lineskiplimit\z@\ialign{$\m@th#1\hfil##\hfil$\crcr#2\crcr\sim\crcr}}}
\catcode`@=12 

\begin{document}
\begin{titlepage}
\begin{flushright}
        {CERN--TH/96--205}
\end{flushright}
\vspace{\fill}\vspace{\fill}
\centerline{\LARGE\bf\strut Transverse mass as a means}
\centerline{\LARGE\bf\strut of measuring the $W$ width}
\centerline{\LARGE\bf\strut at the Tevatron}
\vspace{4ex}
\centerline{\large\bf David Summers}
\centerline{TH Division, CERN}
\centerline{CH-1211 Geneva 23}
\vspace{\fill}\vspace{\fill}
 
\centerline{\large\bf Abstract}
\begin{quote}
At the Tevatron the transverse mass is used to separate on mass shell
from off mass shell $W$ production; and the rate of off
mass shell $W$ production gives a measure of the $W$ width. We look at
alternative variables to see if the separation of on and off mass
shell $W$'s can be improved, and hence give a better measure of the
$W$ width. We find that the transverse mass is very close to the
optimal variable for separating on from off mass shell $W$ decay, and
hence there is little to be gained by using other, more complicated,
variables. This happens because if the transverse mass is above the
$W$ mass, the $W$ is guaranteed to be produced off mass shell.

\end{quote}
 
\vspace{\fill}\vspace{\fill}
 
\begin{flushleft}
CERN--TH/96--205\\
August 1996
\end{flushleft}
\end{titlepage}

In the Standard Model of particle physics there are three massive vector
bosons, the $Z^0$ and the $W^\pm$ bosons. As these bosons are
massive they have finite lifetimes and decay. The $Z^0$ boson, being
electrically neutral, can be produced cleanly in $e^+ e^-$
annihilation. This means that its properties can be accurately
measured. On the other hand the $W$ bosons are electrically charged, and
so cannot be produced in isolation in $e^+ e^-$ annihilations; instead
currently real $W$ bosons are produced in $p \bar p$ colliders, where
the incoming partons can have net charge $\pm 1$. However 
$p \bar p$ colliders are not a clean environment in which to observe
the $W$ decay, where the decay via jets is typically hidden
behind large QCD backgrounds. This leaves only the leptonic decay of
the $W$ to be observed; however, again because the $W$ boson is
charged, leptonic decays always involve an electrically neutral
neutrino which goes undetected. Hence, although the properties of the
$Z$ boson and its decay are accurately measured, the properties of the
$W$ bosons and its decays are far less well known.

For the case of the $W$ width no direct measurement can be made,
instead there are currently two indirect methods of measuring the $W$
width. In the first the ratio of
dilepton $Z$ events at the Tevatron is compared to single lepton +
missing transverse 
energy $W$ events \cite{indirectW}. We have
\begin{eqnarray}
{\sigma(pp \to W \to l\nu) \over \sigma(pp \to Z \to ll) }
&=& {\sigma(pp \to W) \over \sigma(pp \to Z ) }
 {{\rm Br}(W \to l \nu ) \over {\rm Br}(Z \to l l )} \nonumber\\
&=& {\sigma(pp \to W) \over \sigma(pp \to Z ) }
  {\Gamma_Z \over \Gamma_W }
 {\Gamma(W \to l \nu ) \over \Gamma(Z \to l l )}
.
\end{eqnarray}
Now ${\sigma(pp \to W) \over \sigma(pp \to Z ) }$ and 
${\Gamma(W \to l \nu ) \over \Gamma(Z \to l l )}$ can be well predicted
within perturbation theory; $\Gamma_Z$ is accurately measured at LEP,
and so this gives a measurement of $\Gamma_W$. Of course this assumes
that ${\sigma(pp \to W) \over \sigma(pp \to Z ) }$ and 
${\Gamma(W \to l \nu ) \over \Gamma(Z \to l l )}$ can be accurately
predicted, which is in turn based upon assumptions such as that
physics beyond the Standard Model does not modify these quantities.
 
In the second method the shape of the transverse mass $M_T$, 
spectrum of isolated lepton + missing energy $W$ events, is measured
\cite{WwidthTH,CDF,UA1,UA2}, where the transverse mass is defined by
\begin{equation}
M_T^2  = 2 E_{T\nu} E_{Tl} - 2 {\bf p}_{T\nu} \cdot {\bf p}_{Tl}. 
\end{equation}
Now the transverse mass is always less than the actual mass, \ie
\begin{equation}
p_W^2 \ge M_T^2 ;
\end{equation}
so, if the transverse mass is larger than $M_W$ then the
intermediate $W$ must have been forced above its mass shell. Whereas on
shell intermediate $W$ bosons feel the effect of the $W$ width in the
Breit-Wigner propagator, for off shell ones the
width term in the Breit-Wigner propagator is dominated by the standard
term. This means that the rate of these off shell intermediate $W$
bosons is proportional to the $W$ width;
hence the normalisation of the tail of the
transverse mass distribution is sensitive to the $W$ width
$\Gamma_W$. This assumes that the leptonic decay of processes that
take place via an off shell 
intermediate $W$ can be related to those that proceed via an
on shell intermediate $W$.

In this paper we study the second method, to see if it can be
improved to give a more accurate determination of the $W$ width. 
The crux of this method is the ability to separate on shell $W$
production from processes where an off shell intermediate $W$ is
exchanged, and it is not immediately clear that $M_T$ is the optimal
variable to make such a separation. We have
\begin{eqnarray}\label{pw2}
p_W^2  &=& E_{T\nu} E_{Tl}  
    \left( \exp (\Delta \eta) + \exp ( - \Delta \eta ) \right)
    - 2 {\bf p}_{T\nu} \cdot {\bf p}_{Tl} \nonumber \\
&=& M_T^2 + E_{T\nu} E_{Tl}
   \left( \exp (\Delta \eta) + \exp ( - \Delta \eta ) - 2 \right),
\end{eqnarray}
where $\Delta\eta = | \eta_l - \eta_\nu |$. Now as the neutrino
rapidity is unmeasured we cannot evaluate $\Delta\eta$ and hence not
exactly reconstruct $p_W^2$. When the
unobserved neutrino has the same rapidity as the final state lepton
$M_T^2$ equals $p_W^2$; as this is a vanishingly small part of phase
space this clearly never exactly happens. Indeed it is not even clear,
as lepton and transverse neutrino momenta vary from event to event, that
$M_T^2$ has approximately the same behaviour as $p_W^2$. As both $M_T^2$
and $p_W^2$ have the same dependence on $E_{T\nu}$ and $E_{Tl}$, and
${\bf p}_{T\nu}$ and ${\bf p}_{Tl}$ are usually almost back to back, we
shall concentrate on how $M_T$ can be improved by measurements of the
only remaining observed variable, the lepton rapidity $\eta_l$.

As $M_T$ is relatively safe with respect to higher order corrections we
shall use a leading order Monte Carlo simulation of $W$ production at
the Tevatron, that is a $p \bar p$ collider with $\sqrt s = 1.8 \TeV$;
for the $W$ propagator we use a Breit-Wigner propagator
with a running width term
\begin{equation}
\mbox{Prop}_W = { 1 \over (p_W^2-M_W^2) ^2 + p_W^4 \Gamma_W^2 / M_W^2 }
\end{equation}
which Dyson sums the imaginary part of the vacuum polarisation
contribution via massless fermions to the $W$ propagator.
We replace the coupling in the decay of the $W$ boson into leptons by
the $W$ width,
\begin{equation}
g_W^2={\Gamma_W^{\rm leptonic} \over 6\pi M_W}
     ={\Gamma_W \mbox{Br}^{\rm TH}(W\to l\nu) \over 6\pi M_W}
\end{equation}
where $\mbox{Br}^{\rm TH}(W\to l\nu) 
= \Gamma_W^{\rm TH-leptonic} / 
   \Gamma_W^{\rm TH-total}
= 1/(9 + 6 \alpha_S(M_W^2)/\pi)
= 0.10810$.
Replacing the decay coupling
constant ensures that in the narrow width approximation the
overall cross-section to produce a $W$ is independent of the $W$ width
$\Gamma_W$, as we physically expect (due to the physical independence
of production and decay of a $W$). Notice that the use of
$\mbox{Br}^{\rm TH}$ rather than the unmeasured experimental value
changes only the overall number of the $W$ events; it does not
change the shape of $W$ distributions.

For the parton distributions we use MRS D0$'$ \cite{MRS} evaluated at a
scale of $\sqrt{p_W^2}$, which at the Bjorken $x$
and $Q^2$ values probed in $W$
production at the Tevatron should give accurate results. For the
remaining physical parameters we use the tree level Standard Model
values with,
\begin{eqnarray}
M_W &=& 80.22 \GeV \\
\alpha &=& 1/128 \\
\sin^2 \theta_w &=& 0.23 .
\end{eqnarray}
Experimentally, measurements of $W$ bosons at the Tevatron have large
errors due to the unobserved neutrino in the leptonic decay; we model
these errors by giving the measured missing transverse energy the
normal distribution
\begin{equation}
{\cal P}(\slE) = {\exp(-(\slE-E_{T\nu})^2/ (2\GeV) E_{T\nu}) 
                  \over \sqrt{2 \pi E_{T\nu}/(\mbox{GeV})}} ,
\end{equation}
relative to the transverse neutrino energy.
Although this form of smearing is vastly simpler than the actual
experimentally measured smearing of observables, it takes the dominant
smearing into account.

We assume that the distribution of the smearing is known exactly; in
practice we expect that this will be measured accurately in other
processes such as $Z$ decay, to the extent that it will only have a
minimal effect on the $W$ measurements.

As $W^+$ can be experimentally distinguished from $W^-$ from the
charge of the lepton with which it decays, and the $W^-$ distributions
are identical to the $W^+$ distributions under the transformation
$\cos\theta\leftrightarrow -\cos\theta$, we plot $W^-$ events
reversing the sign of $\cos\theta$. This means that we gain extra
information from the forward/backward asymmetry of the charged
leptons, which would be symmetrised away if we did not distinguish
the charges of the leptons.

We shall apply the same cuts on transverse energy as
CDF use, \ie
\begin{equation}\label{cuts}
E_{Tl} > 30 \GeV \qquad \slE_T > 30 \GeV .
\end{equation}
We do not apply the CDF cut on
the lepton rapidity ($|\eta_l|<1.05$) as we retain this as a variable in
all calculations \cite{CDF}. 

We use two methods of measuring how capable different variables are of
measuring the $W$ width. All variables that we consider are sensitive
to the $W$ width in some regions where off mass shell $W$ production
dominates, and insensitive in other regions where on mass shell $W$
production dominates. We estimate the
region where off shell $W$ production dominates by the region where
the variable shows half or more of the full dependence on the $W$ width;
we then use the cross-section in this region as a measure of the ability
of this variable to measure the $W$ width.
Now the cross-section in this region can be measured experimentally
with an accuracy equal to the square root of the number of events in that
region. We therefore expect to be able to measure the $W$ width
with an accuracy:

\begin{equation}\label{error}
{\Delta\Gamma_W \over \Gamma} \simeq 
{1 \over \sqrt{\sigma_{\rm off-shell}\int{\cal L}}}.
\end{equation}

In the second method we generate 22739 unweighted $W$ events that pass
the cuts (\ref{cuts}) with, 
\begin{equation}
\Gamma_W=\Gamma_W^{\rm TH}={\alpha_{\rm em} M_W \over 12 \sin^2\theta_W}
(9+6\alpha_S(M_W^2)) =2.10\GeV;
\end{equation}
this corresponds to about $\int{\cal L} = 20 \mbox{ pb}^{-1}$.
We then perform a binned log-likelihood fit for various values of the
$W$ width. 
The maximum of the log-likelihood gives the central
estimate for the $W$ width, and the region in which the log-likelihood
drops by 0.5 gives the 1 standard deviation error (assuming that the
errors are normal, or equivalently that the log-likelihood is
parabolic). We always plot the log-likelihood relative to the maximum
log-likelihood, as the absolute scale contains no physical
information, depending on factors such as the bin width. Just as CDF,
we allow the normalisation of the differential cross-section to float,
and only extract the log-likelihood from the shape of the differential
cross-section. However, unlike CDF, we calculate the log-likelihood from
the full differential cross-section, rather than just in the region
where this is not sensitive to the experimental
mismeasurement errors; as for our model, we know the effect of the
experimental smearing exactly. This also saves considerable effort
evaluating the region where we are insensitive to the experimental
smearing for each variable that we consider.

We first compare the standard cross-section differential with respect
to the transverse mass $M_T$ with the best possible differential
cross-section if the missing neutrino rapidity $\eta_\nu$ were
known, that is the cross-section differential with respect to
$\sqrt{p_W^2}$, where $p_W^2$ is defined by
\begin{equation}
p_W^2  = \slE_T E_{Tl}  
    \left( \exp (\Delta \eta) + \exp ( - \Delta \eta ) \right)
    - 2 \;\mbox{/}\hskip -0.6em {\bf p}_T \cdot {\bf p}_{Tl}
\end{equation}
in Eq.~(\protect\ref{pw2}) with
$E_{T\nu}$ replaced by $\slE_T$.

\begin{figure}[t]
\epsfxsize=0.8\hsize\epsffile{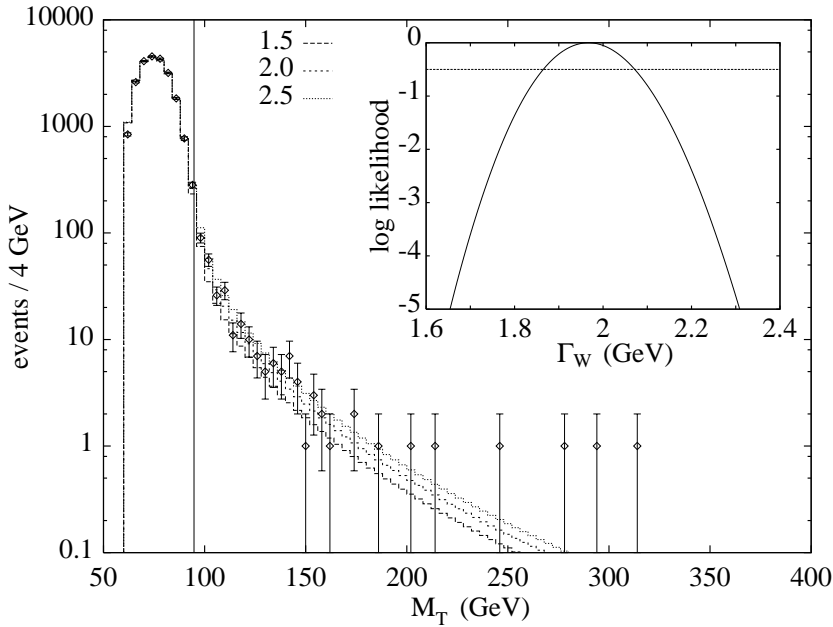}
\caption[]{
$d\sigma/dM_T$ : for a sample of 22739 simulated events with 
$\Gamma_W=2.10\GeV$; the
3 theoretical curves are for $\Gamma_W = 1.5,2.0,2.5 \GeV$. Inset is
the relative log-likelihood for different values of the $W$ width.
}
\protect\label{mt} 
\end{figure}

\begin{figure}[t]
\epsfxsize=0.8\hsize\epsffile{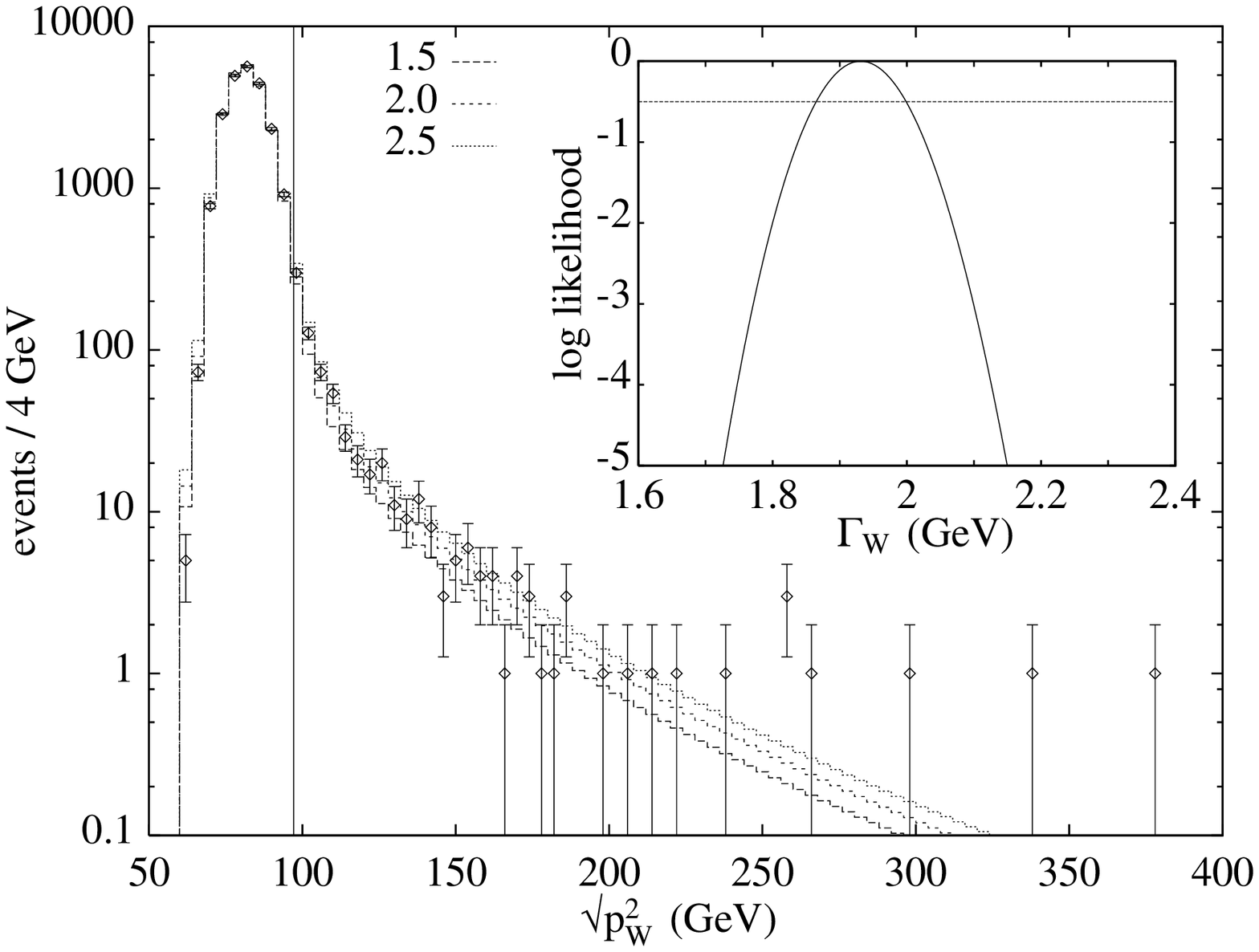}
\caption[]{
$d\sigma/d\sqrt{p_W^2}$ : for a sample of 22739 simulated events with 
$\Gamma_W=2.10\GeV$; the
3 theoretical curves are for $\Gamma_W = 1.5,2.0,2.5 \GeV$. Inset is
the relative log-likelihood for different values of the $W$ width.
}
\protect\label{perf} 

\end{figure}

In Figs.~\ref{mt} and \ref{perf} we plot the event rate
for the unweighted $W$ events vs. the transverse mass $M_T$ and the
estimated momentum flowing through the $W$, $\sqrt{p_W^2}$ respectively. We
also show 
the theoretical differential cross-section, normalised to the same
number of events as the unweighted events, for the 3 values of the $W$
width,
\begin{equation}
\Gamma_W = 1.5 \qquad , \qquad 2.0 \qquad , \qquad 2.5 \GeV .
\end{equation}
We show the dividing line, where the distribution shows half the full
dependence on the $W$ width. In the inset graph we show the
log-likelihood vs. the $W$ width. This gives $1\sigma$ errors in the
measurement of the $W$ width:
\begin{eqnarray}
\Delta\Gamma_W^{M_T} &=& 0.103 \GeV \\
\Delta\Gamma_W^{{p_W^2}} &=& 0.067 \GeV .
\end{eqnarray}
The cross-sections in the off-shell
regions are given by
\begin{eqnarray}
\sigma_{\rm off-shell}^{M_T} &=& 13.5 \pb \\
\sigma_{\rm off-shell}^{{p_W^2}} &=& 24.2 \pb .
\end{eqnarray}
For the transverse mass case the off-shell cross-section and $1\sigma$
errors agree well with Eq.~(\ref{error}); however, for the $W$
invariant mass case the off-shell cross-section gives an error 50\%
larger than the log-likelihood error. This is because the
log-likelihood also gains some sensitivity to the $W$ width from the
region $p_W^2 < M_W^2$, \ie from $W$'s that are produced below
mass shell; whereas the off-shell cross-section comes just from the
region where the $W$ is above its mass shell. This also explains why
the $W$ width extracted from the log-likelihood fit of the
$\sqrt{p_W^2}$ spectrum is considerably lower than extracted from the
$M_T$ distribution. There are relatively few events with low 
$p_W^2$,  as $\sqrt{p_W^2}$ is sensitive to the $W$ width in this
region this drags the extracted width down. Whereas
although there are also relatively few events at low $M_T$, because the
low $M_T$ region is not sensitive to the $W$ width, this does not pull
the extracted $W$ width down. Nevertheless it is
clear that the $\sqrt{p_W^2}$ spectrum gives at least 30\% improvement
in the $W$ width measurement. In this paper we see if this improvement
can be accessed.
\begin{figure}[t]

\epsfxsize=0.8\hsize\epsffile{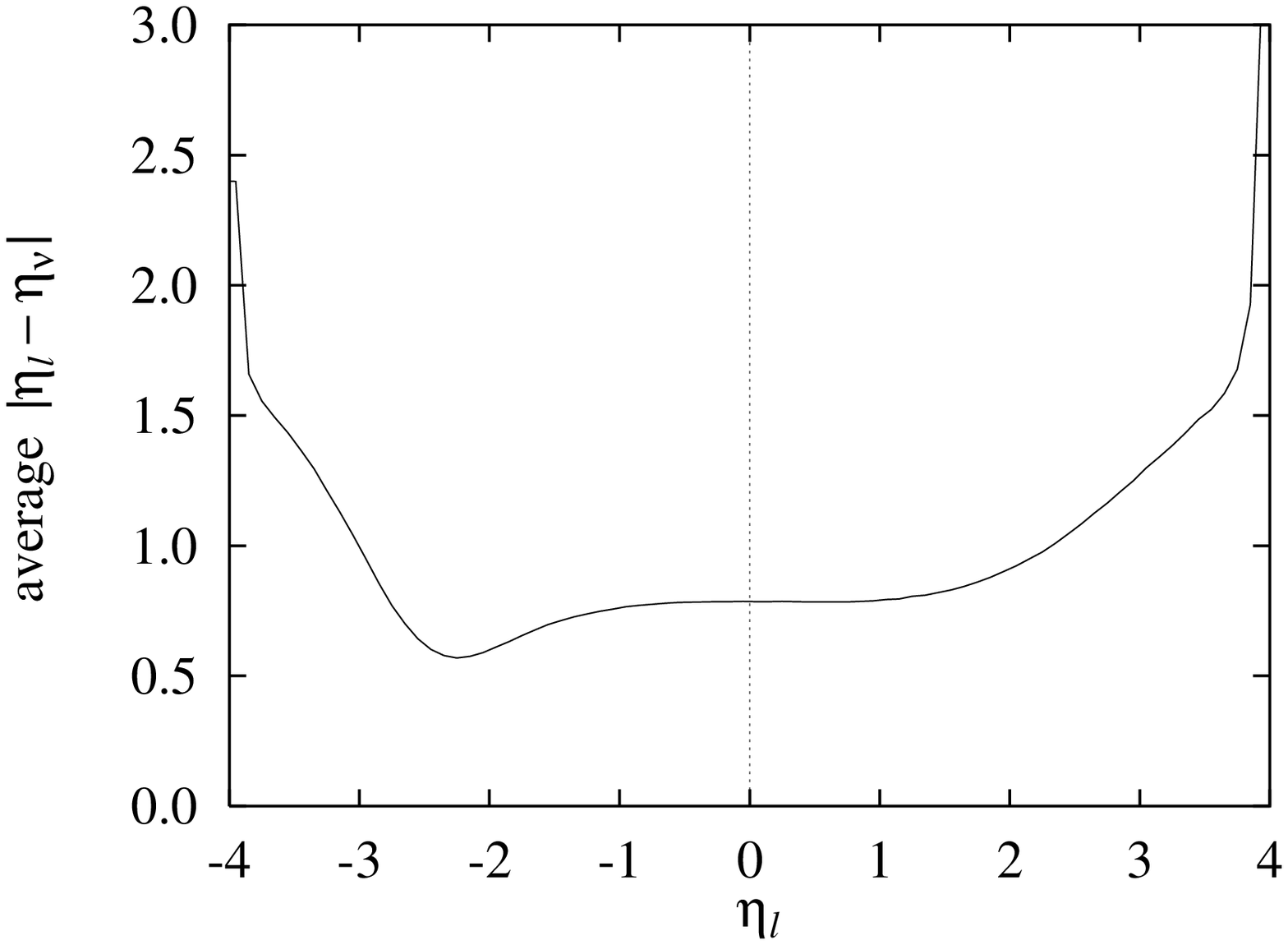}
\caption[]{
The average value of $|\eta_l-\eta_\nu |$ vs. $\eta_l$ for $W$ events.
}
\protect\label{avdeta} 
\end{figure}

Returning to Eq.~(\ref{pw2}) the first question to ask is how the
average rapidity difference between the lepton and the neutrino varies
as a function of the lepton rapidity. Naively we expect $W$'s to be
produced fairly centrally, so if the lepton is produced at large
rapidities we expect the neutrino to be in the opposite hemisphere and
$|\eta_l-\eta_\nu |$ to be large. This can be seen in the actual
distribution, which is plotted in Fig.~\ref{avdeta}. Notice that with
the cuts 
(\ref{cuts}) an on-shell $W$ can produce a lepton with maximum
rapidity 3.91, with $|\eta_l-\eta_\nu |=1.09$; as $p_W^2$ grows larger
than $M_W^2$, the maximum lepton rapidity grows slowly to 4.09, while
the associated $|\eta_l-\eta_\nu |$ grows rapidly, as can be seen in
Fig.~\ref{avdeta}.

For larger values of $|\eta_l|$, $|\eta_l-\eta_\nu |$ grows rapidly,
which tells us that, for large values of $|\eta_l|$, the transverse
mass significantly underestimates the momentum flowing through the
$W$, especially in comparison to small values of $|\eta_l|$. This
suggests using
\begin{equation}
p_{W,\rm est}^2 = M_T^2 + \slE_T E_{Tl}
   \left( \exp (\overline{\Delta \eta}(\eta_l)) 
               + \exp ( - \overline{\Delta \eta}(\eta_l) ) - 2 \right),
\end{equation}
where $\overline{\Delta \eta}(\eta_l)$ is the average value of 
$|\eta_l-\eta_\nu |$ shown in Fig.~\ref{avdeta}.

\begin{figure}[t]

\epsfxsize=0.8\hsize\epsffile{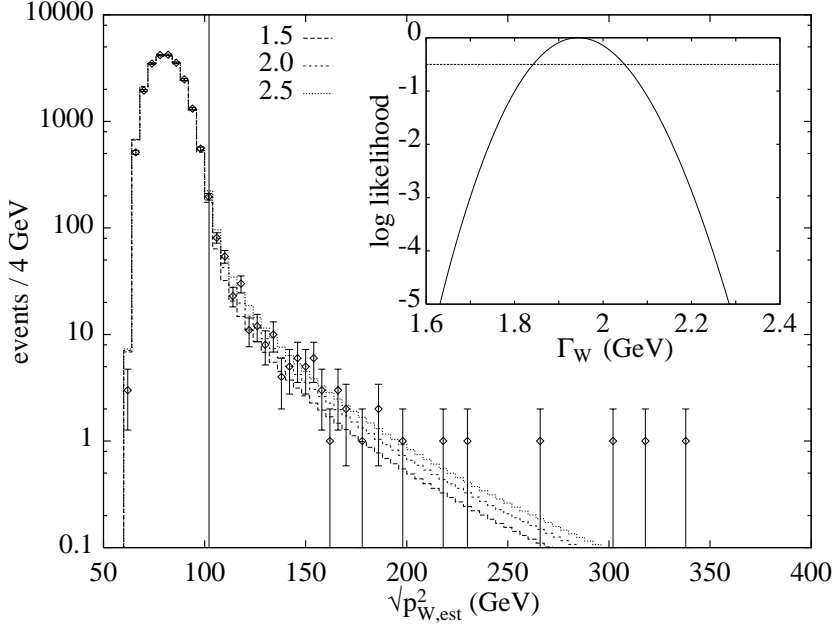}
\caption[]{
$d\sigma/d\sqrt{p_{W,\rm est}^2}$ : 
for a sample of 22739 simulated events with 
$\Gamma_W=2.10\GeV$; the
3 theoretical curves are for $\Gamma_W = 1.5,2.0,2.5 \GeV$. Inset is
the relative log-likelihood for different values of the $W$ width.
}
\protect\label{alt_r2} 
\end{figure}

We show the analogous graph to Figs.~\ref{mt}~and~\ref{perf} for the variable
$\sqrt{p_{W,\rm est}^2}$ in Fig.~\ref{alt_r2}. The log-likelihood gives
a measurement of the statistical error in the $W$ width of
\begin{equation}
\Delta\Gamma_W^{{p_{W,\rm est}^2}} = 0.103 \GeV .
\end{equation}
The cross-section in the off-shell
regions is given by
\begin{equation}
\sigma_{\rm off-shell}^{{p_{W,\rm est}^2}} = 13.3 \pb .
\end{equation}
Clearly $\sqrt{p_{W,\rm est}^2}$ is no better a variable than $M_T$ in
separating on- from off-shell $W$ production, and it has given almost
identical errors. This is because although 
$\sqrt{p_{W,\rm est}^2}$ is a far better estimator of the  momentum
flowing through the $W$ (it clearly peaks far closer to the $W$ mass
than $M_T$), the region where we are sensitive to the $W$ width
clearly moves up to a far higher value. For $M_T$ values greater
than 95~GeV, the
differential cross-section shows a more than 50\% dependence on the
$W$ width, whereas for $\sqrt{p_{W,\rm est}^2}$ values greater than
102~GeV are needed until we have a more than 50\% dependence on the $W$
width. If there is no experimental smearing, then 
$M_T^2 \le p_W^2$ guarantees that the exchanged $W$ is off mass shell
if the transverse mass is greater than $M_W$; whereas 
$p_{W,\rm est}^2 \simeq p_W^2$ and that on-mass shell $W$ production
dominates off-mass shell production, means that for $p_{W,\rm est}^2$
values just above the $W$ mass squared are most likely to be from
on mass shell $W$ production where $p_{W,\rm est}^2$ overestimates
the momentum flowing through the $W$. To counteract this effect,
$p_{W,\rm est}^2$ has to be larger than $M_T^2$ before the
cross-section becomes sensitive to the $W$ width.

\begin{figure}[t]
\epsfxsize=\hsize\epsffile{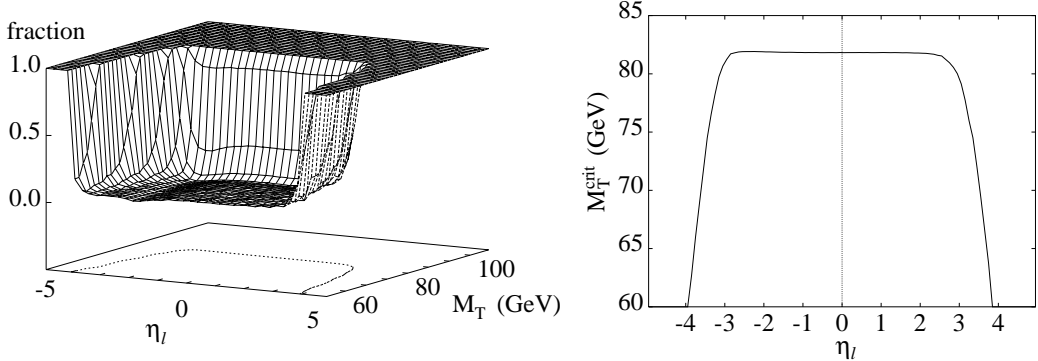}
\caption[]{
The fraction of $W$ events that have $\sqrt{p_W^2} > M_W + \Gamma_W$
vs. $M_T$ and $\eta_l$.
}
\protect\label{offshellvseta} 
\end{figure}

Clearly we should not be interested in a more accurate determination of
$p_W^2$, as the unknown neutrino rapidity means that we can only
reconstruct the $p_W^2$ of an ensemble of $W$ decays; in order
to measure the $W$ width we need to evaluate $p_W^2$ on an event by
event basis. With this in mind, we ask a different question than what
is the average value of $|\eta_l-\eta_\nu |$ for different $\eta_l$
values; what we are more interested in is how the sensitivity to the
$W$ width varies with the lepton rapidity. To be sensitive to the
$W$ width we need the $W$ to be off mass shell, that is 
$\sqrt{p_W^2}-M_W \gsim \Gamma_W$. In Fig.~\ref{offshellvseta} we show
the fraction of events that have $\sqrt{p_W^2} > M_W + \Gamma_W$
as a function both of $\eta_l$ and $M_T$; we also show the $M_T$
required for each $\eta_l$, such that 50\% of the events have 
$\sqrt{p_W^2} > M_W + \Gamma_W$.

\begin{figure}[t]
\epsfxsize=0.8\hsize\epsffile{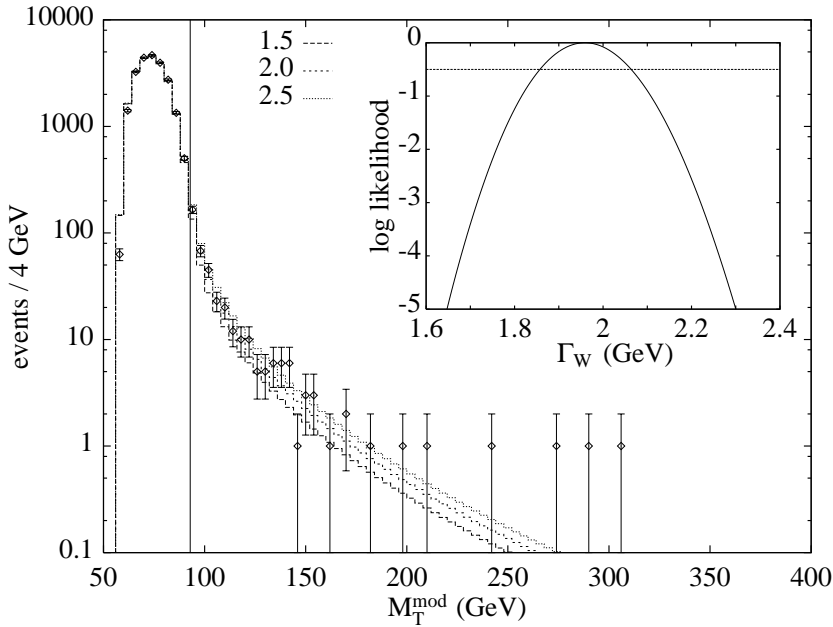}
\caption[]{
$d\sigma/dM_T^{\rm mod}$ : for a sample of 22739 simulated events with 
$\Gamma_W=2.10\GeV$; the
3 theoretical curves are for $\Gamma_W = 1.5,2.0,2.5 \GeV$. Inset is
the relative log-likelihood for different values of the $W$ width.
}
\protect\label{alt_r} 
\end{figure}

What we require is a variable such that for all values of the lepton
rapidity the variable becomes sensitive to the $W$ width at the same
value. From Fig.~\ref{offshellvseta} it is clear that this does not
happen for the transverse mass, although in the central region, with
$|\eta_l|<2.5$, the transverse mass becomes sensitive to the $W$ width
at approximately the constant value of just above the $W$ mass. A
simple modification of $M_T$ that has the property of becoming
sensitive to the $W$ width for the same value, independent of the
lepton rapidity, is
\begin{equation}
M_T^{\rm mod} = M_T { M_W \over M_T^{\rm crit}(\eta_l)}
\end{equation}
where $M_T^{\rm crit}$ is obtained from Fig.~\ref{offshellvseta}. We
scale by $M_W$ in the numerator, so $M_T^{\rm mod}\approx M_T$ for the
central region. We show the analogous graph to
Figs.~\ref{mt}~and~\ref{perf} for the variable 
$M_T^{\rm mod}$ in Fig.~\ref{alt_r}. The log-likelihood gives
a measurement of the statistical error in the $W$ width of
\begin{equation}
\Delta\Gamma_W^{M_T^{\rm mod}} = 0.103 \GeV .
\end{equation}
The cross-section in the off-shell
regions is given by
\begin{equation}
\sigma_{\rm off-shell}^{M_T^{\rm mod}} = 13.5 \pb .
\end{equation}
It can be seen that $M_T^{\rm mod}$ does not lead to a significant
improvement in the measurement of the $W$ width. This happens
because $M_T^{\rm mod}$ is essentially identical to $M_T$ for 
$|\eta_l| < 2.5$, and the vast majority of $W$'s are produced with
$|\eta_l| < 2.5$. This also means that if there is an experimental cut
on the lepton rapidity, such as $|\eta_l|<1.05$ that CDF apply, then
$M_T$ is effectively the optimal variable for separating on- from 
off-shell $W$ decays; certainly there is little to be gained from
measurements of the lepton rapidity.

\section*{Conclusions}

In this paper we look at improvements that can be made to the
transverse mass variable at the Tevatron to enhance measurements of the
$W$ width. In particular we look at enhancements that come from using
the rapidity of the measured charged lepton that comes from the
$W$ decay. We construct 2 new variables, $p_{W,\rm est}^2$, which
estimates the momentum flowing through the $W$, and $M_T^{\rm mod}$,
which for all charged lepton rapidity values becomes sensitive to the
$W$ width at the same value. $p_{W,\rm est}^2$ does not enhance
measurements of the $W$ width, as the region where this variable
becomes sensitive to the $W$ width is moved to higher values, where the
cross-section is lower.

On the other hand, $M_T^{\rm mod}$ is constructed to be the optimal
modification of the transverse mass from measuring the charged lepton
rapidity. However, for values of $|\eta_l|<2.5$, $M_T^{\rm mod}$ and 
the transverse mass are effectively equivalent. Typically there are
very few events with $|\eta_l|>2.5$; this means that 
$M_T^{\rm mod}$ leads to no significant improvement in the measurement
of the $W$ width. This is especially true if there is an experimental
cut on the charged lepton rapidity.

\section*{Acknowledgements}

I would like to thank Mike Seymour for helpful comments on the manuscript.


\begin{thebibliography}{99.}

\bibitem{indirectW} 
CDF Collaboration, F.~Abe et al., Phys.~Rev.~Lett.\ {\bf 73} (1994) 220.

\bibitem{WwidthTH} J.~Smith, W.L.~van~Neerven and J.A.M.~Vermaseren
Phys.~Rev.~Lett.\ {\bf 50} (1983) 1738.

\bibitem{CDF}
CDF Collaboration, F.~Abe et al., Phys.~Rev.~Lett.\ {\bf 74} (1995) 342.

\bibitem{UA1} 
C.~Albajar et al., Z.~Phys.\ {\bf C44} (1989) 15.

\bibitem{UA2}
R.~Ansari et al., Phys.~Lett.~B\ {\bf 186} (1987) 440.

\bibitem{MRS}
A.~D.~Martin, R.~G.~Roberts and W.~J.~Stirling, Phys. Lett. 
{\bf B306} (1993) 145.

\end{thebibliography}
\end{document}